\documentclass[11pt]{article} 

\usepackage[utf8]{inputenc} 

\usepackage{geometry} 
\usepackage{graphicx} 
\usepackage{hyperref} 

\usepackage{booktabs} 
\usepackage{array} 
\usepackage{paralist} 
\usepackage{verbatim} 
\usepackage{subfig} 

\title{ASTOR: Evolutionary Automatic Software Repair for Java}
\author{Matias Martinez, Martin Monperrus}
\date{Technical Report hal-01075976, Inria, 2014}

\begin{document}
\maketitle

\begin{abstract}
\textbf{Context:}
During last years, many automatic software repair approaches have been presented by the software engineering research community. 
According to the corresponding papers, these approaches are able to repair real defects from open source projects.

\textbf{Problematic:}
Some previous publications in the automatic repair field do not provide the implementation of theirs approaches.
Consequently, it is not possible for the research community to re-execute the original evaluation, to set up new evaluations (for example, to evaluate the  performance against new defects) or to compare approaches against each others.

\textbf{Solution:}
We propose a publicly available automatic software repair tool called Astor.
It implements three state-of-the-art automatic software repair approaches in the context of Java programs (including GenProg and a subset of PAR's templates).
The source code of Astor is licensed under the GNU General Public Licence (GPL v2).
\end{abstract}

\section{Introduction}
During last years, many automatic software repair approaches have been presented by the software engineering research community. 
According to the corresponding papers, these approaches are able to repair real defects from open source projects.
Some previous publications in the automatic repair field do not provide the implementation of theirs approaches of the evaluation details  \cite{Monperrus2014}.
Consequently, it is not possible for the research community to re-execute the original evaluation, to set up new evaluations (for example, to evaluate the  performance against new defects) or to compare approaches against each others.
Consequently, we propose a publicly available automatic software repair tool called Astor.

Astor (\underline{A}utomatic \underline{S}oftware \underline{T}ransformations f\underline{O}r program \underline{R}epair) is a tool for automatically repairing Java programs. It is a test-suite based repair tool:
Astor takes as input the source code of a buggy program and its test suite, with at least one failing test case.

Along the same line as Genprog \cite{Weimer2009}, it uses a kind of evolutionary computation. Astor's main goal is to evolve a software program until a given goal, i.e., \emph{objective goal} is reached (usually ``all test cases pass'').

Astor is not only a specific repair tool but also a generic a framework for implementing repair approaches that use common components such as fault localization.
For instance,
we use the generic Astor framework to implement three state of the art automatic software repair approaches:
GenProg \cite{Weimer2009} (for Java and not for C),
a subset of PAR's templates  \cite{Kim2013}, 
and a mutation-testing based repair approach \cite{debroy2010using}.

\section{How Astor Works?}
Astor uses genetic programming \cite{koza1992genetic} as main evolutionary paradigm.
In the automatic software repair field, previous works \cite{Weimer2009, ArcuriEvolutionary} have also used genetic programming for searching candidate patches.

\subsection{Foundations of Astor: Genetic Programming}
Astor takes as input a program to evolve.
The output is a set of programs that fulfill a given criterion.
First, Astor creates a population of $n$ programs, that are all clones of the input program. 
Each one is called a \emph{program variant} $Pv_i$. At this point, all those variants are identical.

Then, Astor evolves those programs over a fixed number of  iterations (in Genetic Programming an iteration is know as a generation). 
In Astor, a program variant of generation $n$ produces one or more child program variants for the next generation $n+1$.

On each generation,
Astor applies one source code transformation $T$ in each program variant $Pv_i$ (the \emph{parent variant}). 
The resulting program that combines the parent variant with the transformation result, is called a \emph{child variant}. 
After applying the transformations, Astor evaluates a fitness function $f$ over each $Pv_i$ to know whether new variants reach the predefined goals. 
The output of $f$ usually is a numerical value or a boolean. 
For example, in the automatic software repair context, this fitness function indicates whether a program has a defect or not.
Then, after each child variant is evaluated, 
Astor selects $n$ variants, between the children and parents, to be part of the next generation, according to their fitness values.

Astor finishes the evolution when one of the following  situations happen:
a) It executes $MaxGen$ generations;
or b) It finds a program variant that fulfills the goals, i.e., the fitness value of that variant is equal to a desired value.

\subsection{Astor as an Automatic Repair framework}

In the automatic software repair field, the goal is to automatically find a patch that repairs a defect in the input program.

Astor considers test suite based automatic repair: the test suites are used as a proxy to the program specification  \cite{Goues2012journal}. 
That means, a program is considered as fulfilling its specification if all the test cases composing the test suite pass. Otherwise the program has a defect.

Astor implements three state of the art test suite based repair approaches: GenProg \cite{Weimer2009} (JGenProg in Astor),
a subset of PAR's templates  \cite{Kim2013}, 
and a mutation-testing based repair approach \cite{debroy2010using}.
In those approaches, the fitness relates to the number of failing test cases. 
The goal is to find a program variant with a fitness value of zero i.e., no failing test cases.

Astor takes as input the source code program, at least one failing test case, the number of max generations (iterations) and program variant population size (number of program variant that are modified in each generation).
Astor is built over Spoon \footnote{http://spoon.gforge.inria.fr/}, a library for Java code analysis and manipulation.
Thanks to Spoon, all source code transformations are done at the AST level \cite{Martinez2013}.

\section{Architecture of Astor}

We have used Astor for implementing three state-of-the-art repair approaches.
They are GenProg \cite{Weimer2009}, a subset of PAR's templates \cite{Kim2013} and the mutation-based repair approach from Debroy and Wong \cite{debroy2010using}.
All three are test-suite based repair approaches.
In Astor, there are common components shared across those different repair approaches.
They are:
\begin{enumerate}
\item the fault localization that is used to detect suspicious statements and the order in which these suspicious components are considered to apply a candidate fix;
\item the way a candidate fix is validated.
\end{enumerate}

\subsection{Fault Localization}
\label{sec:fault-localization}

Fault localization consists of computing suspicious statements (statements suspected to contain a bug).
The three considered approaches use fault localization techniques based on spectrum analysis.
Spectrum based fault localization approaches execute test cases of a program and trace the execution of software components (e.g., methods, lines).
Then, formulas calculate the suspicious value of each component. 
These formulas take as input the collected traces and the test results.
The suspicious value goes from 0 (low probability that the statement contains a bug) to 1 (high probability of bugginess).
GenProg and PAR use a formula presented by Weimer et al. in \cite{Weimer2009}, while MutRepair uses the Tarantula formula \cite{Jones2002}.

As the input (program code and test suite) and output (suspicious statement list) of these different techniques do not differ, 
one can change the fault localization technique that is used.
In particular, one can unify all repair techniques so that they use the exact same fault localization technique. 
This is what our framework does. It uses the Ochiai formula \cite{abreu2006evaluation} in the implementation of the three approaches.

Furthermore, the navigation of the fault localization space means to pick one suspicious statement from the space.
This selected statement is modified according to some repair operators.
They are different navigation strategies,
one can select the suspicious statement:
\begin{inparaenum}[\itshape a\upshape)]
\item in suspiciousness order, from the most suspicious statement to the least one;
\item in a uniformly random manner: the suspicious statement are randomly selected before application of the repair operator;
\item with a weighted randomly strategy: the probability to select a suspicious statement is proportional to its suspicious value.
\end{inparaenum}
As for the fault localization technique, our framework unifies this point: all three approaches implement the weighted random strategy.

To sum up, the same fault localization technique and navigation strategy is used in the implementation of the three considered repair approaches.

\subsection{Repair Validation}
\label{sec:repair-validation}
Once a repair operator is applied, one obtains a program variant, called a candidate repair.
This candidate repair then has to be validated against the correctness oracle, the test suite. 
However, there are different ways to do this. 
Our framework does it one way, but the same for repair techniques under consideration. 

First, it executes the originally failing test cases  over the modified program.
If these test cases now pass successfully, it means that the bug is fixed.
However, the patch may have broken an existing functionality. 
To check this, a regression test is executed, it verifies whether the candidate repair breaks anything. The regression test involves executing all previously passing test cases from the test suite.
If none of them fails, the candidate repair is considered valid.
If at least one test fails during the validation phase, the candidate repair is incorrect and is discarded.

\subsection{Extensibility of Astor}

One can define other extension points in the repair framework of Astor.
One can integrate test cases prioritization techniques \cite{rothermel2001prioritizing} to reduce the time to execute regression validations. 
For instance, Qi et al. \cite{mao2013priorization}  presented a prioritization technique  applied in an automatic program repair approach. 
One can also customize the way existing code is reused to form a patch (local versus local strategy) \cite{martinez2014icse}.

\section{Conclusion}
We presented Astor, a evolutionary automatic software repair framework.
Astor fixes bugs in Java code where the bug oracle is a failing test case.
The code is publicly available at \url{https://github.com/SpoonLabs/astor} under the GNU General Public Licence (GPL v2) and is free of use as long as this technical report is properly cited.

\newpage
\bibliographystyle{plain}
\bibliography{biblio-software-repair}

\end{document}